\documentclass[sigconf]{acmart}
\AtBeginDocument{%
  }
    
\usepackage{cleveref}
\usepackage{enumitem}

\newcommand{\etal}[1]{{#1}~et~al.}
\newcommand{\eg}{e.g.,~}
\newcommand{\ie}{i.e.,~}
\newcommand{\cf}{cf.~}
\newcommand{\textQuote}[1]{``\emph{#1}''}

\crefname{section}{Sec.}{Sec.}
\crefname{figure}{Fig.}{Fig.}
\crefname{table}{Tab.}{Tab.}

\setcopyright{none}
\acmYear{2025}
\acmDOI{}
\acmISBN{}
\acmConference[CHI'25 Workshop: Beyond Glasses -- Future Directions for XR Interactions within the Physical World]{1st International Workshop on XR Interactions within
the Physical World (XR-Phy) at CHI 2025}{April 26, 2025}{Yokohama, Japan}

\begin{document}



\title
[Affordance Fallacies in AR]
{Of Affordance Opportunism in AR: Its Fallacies and Discussing Ways Forward}


\author{Marc Satkowski}
\orcid{0000-0002-1952-8302}
\email{msatkowski@acm.org}
\affiliation{%
  \institution{Interactive Media Lab Dresden, \\ Dresden University of Technology}
  \city{Dresden}
  \country{Germany}
}
\author{Weizhou Luo}
\orcid{0000-0002-1312-1528}
\email{weizhou.luo@tu-dresden.de}
\affiliation{%
  \institution{Interactive Media Lab Dresden, \\ Dresden University of Technology}
  \city{Dresden}
  \country{Germany}
}
\author{Rufat Rzayev}
\orcid{0000-0002-0466-2445}
\email{rufat.rzayev@tu-dresden.de}
\affiliation{%
  \institution{Interactive Media Lab Dresden, \\ Dresden University of Technology}
  \city{Dresden}
  \country{Germany}
}

\renewcommand{\shortauthors}{Satkowski et al.}
\acmArticleType{Position}


\begin{abstract}

This position paper addresses the fallacies associated with the improper use of affordances in the opportunistic design of augmented reality (AR) applications. 
While opportunistic design leverages existing physical affordances for content placement and for creating tangible feedback in AR environments, their misuse can lead to confusion, errors, and poor user experiences. 
The paper emphasizes the importance of perceptible affordances and properly mapping virtual controls to appropriate physical features in AR applications by critically reflecting on four fallacies of facilitating affordances, namely,
    the subjectiveness of affordances, 
    affordance imposition and reappropriation, 
    properties and dynamicity of environments,
    and mimicking the real world.
By highlighting these potential pitfalls and proposing a possible path forward, we aim to raise awareness and encourage more deliberate and thoughtful use of affordances in the design of AR applications.


\end{abstract}

\maketitle


\section{Introduction}
\label{ch:intro}

Augmented Reality (AR) glasses enhance how we interact with digital information in our everyday and professional lives. As these devices become increasingly sophisticated and unobtrusive, as seen with devices like
    the Snapchat Spectacles~\cite{spectaclesSpectacles}, 
    the XREAL glasses~\cite{xreal}, 
    or the recently presented Meta Orion~\cite{metaOrionGlasses}, 
they allow digital overlays to seamlessly blend with our physical environment, offering a more straightforward integration, enhanced interaction, and extended use.

The growing accessibility and versatility of AR glasses open up new possibilities for interaction design, particularly in leveraging the physical world as an interface. 
For instance, proxemic interactions~\cite{ballendat2010proxemic,marquardt2012informing} advocate that AR head-mounted displays (HMDs) \textQuote{have fine-grained knowledge of nearby people[,] other devices}~\cite{marquardt2012informing}, and even non-digital physical objects~\cite{ballendat2010proxemic}. 
Since AR inherently allows users to access their surroundings, it is promising to explore interaction techniques that leverage new affordances from the surrounding environment, such as those related to other digital devices, social interactions, or physical objects. 
In general, affordances can be utilized to create a more intuitive interaction design that can rely on the user experience, knowledge, and perception. 
While using the affordances from the environment can help maintain a connection to physical space, their opportunistic application (\cf \cite{henderson2008opportunistic}) can lead to interesting and novel concepts. 
Nonetheless, these concepts may not always be applicable to real-world usage and can pose challenges for interaction design. 
Therefore, we asked ourselves: \textit{How well can an opportunistic use of affordances translate to the future, widespread, real-world usage of AR systems?}

With this position paper, we aim to start a discussion about the meaningful and appropriate use of affordances in AR applications.
While we cannot fathom every possible aspect of this broad topic, we believe that starting this conversation and making future researchers and practitioners aware of potential fallacies is of utmost importance.
To do that, in the following, we
    \textbf{(1)} characterize what the terms affordance and opportunistic design entail (see \cref{ch:background}),
    \textbf{(2)} use our own experience as well as related publications to describe fallacies of affordance use (see \cref{ch:reflection}),
    and, finally, \textbf{(3)} provide approaches to help avoid the fallacy of affordance (see \cref{ch:recommendations}).

\section{About Affordance and Opportunistic Design}
\label{ch:background}

Before we describe and demonstrate what fallacies the idea of affordance and its opportunistic use can introduce, it is necessary to explain what affordance means.
For that, we  
    shortly describe what affordances are, 
    what type of affordances AR systems and devices bring, 
    and lastly, how the real-world environment's affordances can be integrated into AR applications.

\paragraph{General \& Perceived Affordance} 

The term "affordance" was coined by Gibson~\cite{gibson1977theory} in 1977 and was extensively used over the years, resulting in a multitude of connotations~\cite{osiurak2017what}, which are far too numerous (and not needed for this discussion) to depict here.
Initially, it describes the \textQuote{complementarity of [an individual] and the environment}~\cite{gibson1977theory}, where the affordance of the latter \textQuote{offers the [individual], what it provides or furnishes, either for good or ill}~\cite{gibson1977theory}.
Affordances are objective and subjective at the same time~\cite{osiurak2017what}, since they exist without an observer (\ie individual)~\cite{gibson1977theory} while the frame of references (\ie context) is the individual's action capabilities~\cite{osiurak2017what}.
However, when it comes to user interface (UI) design, understanding affordances as existing without any actors present is not particularly useful.
We, therefore, want to highlight the term "perceived affordance", described by Norman~\cite{norman1999affordance}.
Following, \textQuote{an object only affords something when an [individual] capable of using [its] features [perceives] such an object}~\cite{steffen2019framework}.
For example ~\cite{norman1999affordance}, all screens within reaching distance afford touching, but only some can detect it -- presenting the difference between the perceived (\ie touchable) and real affordance.

\paragraph{Affordance of AR} 

Affordance \textQuote{is not premade or tied to the users’ inner schema but occurs in a directed interaction between the human and technology}~\cite{shin2022does}.
In the context of AR as a technology, the affordances get shaped by two fundamental features: the combination of physical and virtual scenes and the ability to break physical laws~\cite{steffen2019framework}.
Based on these, \etal{Steffen}~\cite{steffen2019framework} presented a first set of four affordances, along with several sub-affordances, \eg training~\cite{catenazzi2024mixed}), that AR can provide to users: 
    (1) enhancing positive aspects of, 
    (2) diminishing negative aspects of, 
    (3) recreating aspects of, 
    and (4) creating aspects that do not exist in 
the physical world.
However, with technological advancements, new AR features can appear, and even new affordances, as \textQuote{we have not identified all potential affordances}~\cite{steffen2019framework}.
One such additional affordance can be the immersion and presence provided by an AR system, as conceptualized by \etal{Shin}~\cite{shin2022does}.
Both \textQuote{belong neither to the context/environment nor to the individual, but rather to the relationship between individuals and their perceptions of the environment}~\cite{shin2022does}.

\paragraph{Affordance in AR \& Opportunistic Design} 

As AR combines real-world and virtual scenes into one experience, it can rely on affordance from both environments.
The use of real-world environmental affordances is especially of interest.
For example, many publications explored how virtual content can be placed in relation to the real world environment (\eg \cite{luo2022where,chae2020arphya,lee2024design}) or social interaction (\eg \cite{rzayev2020effectsa}).
The physical haptics of the environment and its objects were also investigated, as seen with the concept of "opportunistic controls"~\cite{henderson2008opportunistic,henderson2010opportunistic}.
It is described as \textQuote{a tangible user interface~\cite{ishii1997tangible} that leverages naturally occurring, tactilely interesting, and otherwise unused affordances}~\cite{henderson2008opportunistic}.
However, for all possible uses of environmental affordances, we always have to consider that AR alters the perception of the real world, mainly through the visual channel.
This, in fact, can impact not only the perceived affordances in the real world but also the affordances of virtual objects, especially those recreating real-world objects (\eg~\cite{zhao2021perception,gagnon2020rolea}).

\section{Identifying the Fallacies}
\label{ch:reflection}

Affordances present a user with interaction opportunities, while opportunistic controls employ already existing but unused affordances.
This combination creates fallacies precisely because it follows the assumption that leveraging existing (and even non-existent) environmental affordances always leads to optimal user experiences.
While these interfaces seek to use nearby physical objects for placing and interacting with virtual content, this approach can introduce several potential issues.
Following, we want to describe four groups of fallacies, which were exhibited through reflection on our own experiences and related works.

\subsection{Subjectiveness of Affordances}
\label{ch:reflection:subjectiveness}

Affordances are subjective by nature~\cite{osiurak2017what}.
This means that every individual, given the same environment, could perceive affordances differently.
For example, a user in an AR document management application (\eg \cite{luo2025documents}) can perceive a whiteboard as a suitable space to place a virtual document, while another user may not consider it appropriate for this purpose because of the presence of important physical notes on the board.
Additionally, a user may lack prior experience to recognize what, for example, a physical project in the environment affords.
Therefore, in opportunistic design, a user might overlook interaction possibilities due to not perceiving the affordances.
However, considering multi-user applications raises an interesting question: How does the context of the environment affect the perceived affordances for collaboration among multiple AR users? 
In our prior study~\cite{luo2022where}, we observed that individuals had more diverse ways of organizing documents and utilizing physical environments for scaffolding. 
However, when two participants collaborated, they made more consistent and frequent use of the physical environment, such as tables and whiteboards, as common grounds for collaborative activities. 
This interesting finding highlights the diversity and subjectiveness of affordances, as well as the characteristics of collective affordance~\cite{weichold2020collective} that emerge when multiple users interact.
Additionally, in multi-user scenarios, individuals may learn from one another and become acquainted with new affordances.


\subsection{Affordance Imposition and Reappropriation}
\label{ch:reflection:reapproprition}

Employing \textQuote{unused affordances}~\cite{henderson2008opportunistic} is at the core of opportunistic control for AR.
However, "unused" can have two meanings, especially as \textQuote{[affordances] exist naturally: they do not have to be visible, known, or desirable}~\cite{norman1999affordance}.
On one hand, individuals may have no association with a particular environmental feature, meaning it is unknown to them.
On the other hand, they may be aware of this affordance but fail to perceive it due to differences in context or other environmental factors.
This leads to two possibilities for facilitating those features, through \textbf{imposition} or \textbf{reappropriation}, which both introduce their own set of challenges.


With \textbf{imposition}, a new action capability is enforced (\ie imposed) on an environmental or object feature.
In Ubi~Edge~\cite{he2023ubi}, the opportunistic use of object edges is proposed, like using a teacup brim to control the color of a virtual object.
While the brim provides distinct haptic feedback (cf.~\cite{ishii1997tangible}, an individual may not recognize it as an interactable feature.
A similar problem arises in TriPad~\cite{dupre2024tripad}, where any given surface can be facilitated as a touch surface.
However, how does the user become aware of this capability and its limits?
In some cases, these challenges can be mitigated by leveraging the metaphors that physical objects convey. 
For instance, using a book metaphor allows users to hold and display virtual content, as well as navigate by turning pages, as this action aligns with the users' understanding of the object's affordance~\cite{luo2024pages, billinghurst2001magicbook}.

With \textbf{reappropiation}, another action capability is mapped onto an existing affordance that is not currently perceived (\ie perceived affordance) but is known to the individual.
While this approach seems promising, it is unpredictable whether every potential user will overlook the "original" affordance.
This, in turn, can lead to unexpected situations, such as occlusion or unintended interaction due to dual-purpose functionality. 
As an example from our own work~\cite{satkowski2022investigating}, a floor can provide spatial affordance for placing a virtual overlay, but that overlay may also occlude real-world signage on the floor. 
In another example~\cite{he2023ubi}, a pan handle can be used opportunistically to control a video player despite its primary affordance being for holding the pan. 
This dual use of the object's affordances could mislead users and even pose risks, such as burns due to interacting using a frying pan.
More challenging situations can arise from opportunistic pairings, particularly when dual functionalities conflict (\eg~\cite{hettiarachchi2016annexing,jain2023ubitouch}). 
For instance, a virtual rotary knob that adjusts the intensity of light in an AR application could be paired with a physical knob intended to change the volume of a real-world radio. 
This kind of opportunistic use of affordance can even lead to conflicting functionalities, increasing cognitive load for users as they may need to relearn interface mappings.

Generally, repurposing everyday items for placement and tangible interaction~\cite{funk2014augmented} can lead to a mismatch between virtual and physical objects. 
Research has also shown that a more significant disparity between the physical and virtual objects can negatively impact the believability of the experience and hinder effective interaction~\cite{simeone2015substitutional, kwon2009effects}, highlighting the challenge of maintaining user immersion and engagement when relying on opportunistic placements.
\subsection{Properties and Dynamicity of Environments}
\label{ch:reflection:environment}

Affordances not only rely on the subjective interpretation (see \cref{ch:reflection:subjectiveness}) but also the situated context created by the environment (cf.~\cite{gibson1977theory}).
While it is possible to repurpose certain aspects of these environments and their objects, such as their shape, for interacting with digital content, their original functions and purpose remain independent of such reappropriations (see \cref{ch:reflection:reapproprition}).
This means, that the object in question could not even be present if its primary purpose is not needed, as its secondary uses often become less significant or even hidden.
Also, the quantity of the same type of object (\eg several teacups for each person in a meeting) can make it challenging to know which object provides the interaction capability needed -- so a user recognizes more perceived affordances than there actually exist~\cite{norman1999affordance}.
Furthermore, it is necessary to think about the state in which a given object can be found.
To stay with the example of a teacup, individuals might not want to interact with a teacup that is dirty or already used by another person.
Likewise, a whiteboard that is already completely filled with writing is no longer perceived as a viable space for pinning documents.

To make it more complicated, both properties mentioned above can change dynamically throughout the use of an AR application.
This means that objects can be added or removed from the environment, the state of objects (\eg writing on a whiteboard) can change, or even the entire environment can get "exchanged" as the individual moves around.
In a prior work~\cite{satkowski2022investigating}, we explored the idea of offloading virtual interfaces onto the ceiling or floor.
However, physical environments partly make already use of both areas, like in a supermarket with advertisements and signs on the ceiling or navigational guides on the floor.
When the individual leaves the supermarket, those areas can become content-free or even disappear entirely, like with the ceiling when one steps outside the building.
Lastly, the way an object or environment is used is not only determined by its form and function but also by the spatial, social, and cultural context in which it exists. 
This inherent inconsistency in usage can create challenges in AR environments, where users rely on familiar (see \cref{ch:reflection:subjectiveness}) physical objects to support digital interactions. 
When affordances shift across different settings, users may struggle to identify suitable objects for interaction, leading to confusion and inefficiencies in opportunistic interface design.

\subsection{Misinterpreting AR as the Real World}
\label{ch:reflection:mimicking}

The ongoing technological advancement of AR enables HMDs to become more capable of presenting virtual scenes that resemble the real world, increasing the immersion affordance of AR devices~\cite{shin2022does}.
Even as immersion is the central selling point of AR (\eg mimicking the real world), new problems linked to affordances can be introduced.
As virtual content gets more "realistic", it will become indistinguishable (at least visually) from the real world.
However, this affects perceived affordance -- individuals could expect that a virtual object will provide the same interaction capabilities and behaviors as its real-world counterpart~\cite{norman1999affordance}.
This, again, can create confusion and reduce trust in the application.
Another problem that can arise is the potential (deliberate and unintentional) misuse of the realism of virtual objects.
For example~\cite{satkowski2022investigating}, a virtual overlay that is not flat on the floor could be perceived as an obstacle, if walked against would afford injury~\cite{gibson1977theory}.
Therefore, individuals will try to avoid the virtual overlay, walking in a different direction (\eg into a supermarket).
To overcome this malicious use (\ie dark pattern), the individuals have to actively resist the perceived affordance and walk through the virtual objects since it has no physical component that really can cause harm.

\section{How to Overcome the Fallacies}
\label{ch:recommendations}

We have described several problems and fallacies with the usage of affordance.
In the following, we sketch key ways to overcome these fallacies, both in the short- and long-term.
However, these approaches are not complete and should be treated as a starting step to make affordances more "safe to use" in future AR systems.

\paragraph{Provide Feedforward}

An appropriate use of affordances can contribute to a feeling of intuitiveness.
While Norman highlighted that affordances should \textQuote{exist naturally: they do not have to be visible, known, or desirable}~\cite{norman1999affordance}, discovering available interactions in mixed-reality environments is inherently challenging.
This difficulty arises from the subjectiveness of affordances (see \cref{ch:reflection:subjectiveness}) and the variability of object affordances across environments (see \cref{ch:reflection:environment}).
Feedforward design, which shows users information about available actions and expected outcomes of such actions before execution~\cite{muresan2023using}, offers a promising approach to address these issues.
It can enhance the discovery of new available interactions in MR applications. 
Moreover, feedforward design could reduce misinterpretation and promote a more consistent perception of affordances across users.
However, designing effective feedforward remains challenging. 
It requires balancing information density and timing to avoid disrupting user workflows, overloading cognitive capacity, or diminishing immersion.
We encourage future research on feedforward design for interaction through physical environments.


\paragraph{Design Interaction with Context}

An arbitrary coupling between available interaction and everyday objects can be detrimental or even misleading.
For instance, using the brim of a teacup as a color picker for a virtual object~\cite{he2023ubi} has no semantic (\cf~\cite{ellenberg2023spatiality}) or contextual relation, making the affordance challenging to perceive.
In contrast, placing a color picker on the base's edge of a lamp for the inserted light bulb is far easier to conceive~\cite{he2023ubi}.
Therefore, we caution against the temptation to design affordances based solely on physical objects and their surroundings. 
Instead, design should center on the users, ensuring interactions are contextualized between users and environments while preserving the necessary semantic relations in between~\cite{ellenberg2023spatiality}.
This offers instructive cues, such as referencing the interaction target and clarifying expected outcomes, further enhancing usability. 


\paragraph{Call for Design Methods and Guidelines}

While the importance of affordances in AR interface design has been widely acknowledged, there is a lack of methodological guidance to support the design process.
For instance, how can designers effectively explore the solution space and discover new affordances?
How can they conceptually navigate through numerous design dimensions and parameters offered by physical surroundings?
Must physical objects and environments always be incorporated, or should AR interfaces move beyond replicating the real world (see \cref{ch:reflection:mimicking})?
How can designers use collective affordance~\cite{weichold2020collective} in the design of multi-user AR applications? 
Beyond these methodological gaps, no comprehensive design guidelines exist for affordance-based AR interactions. 
It would be invaluable for designers and practitioners to have a structural overview of design elements, particularly with highlights of which affordances have been tested, their effectiveness in specific contexts and tasks, and common pitfalls. 


\paragraph{Call for Long-term Field Studies}

Affordances emerge from the relations between users and their environments.
Therefore, we see the need for more empirical insights to account for user differences and dynamicity. 
How do affordances vary based on the social, cultural, or personal background?
How do perceptions differ between single- and multi-user scenarios?
More importantly, as affordances are deeply rooted in the context, we want to highlight the necessity of evaluation beyond one-time laboratory studies.
Long-term studies are particularly valuable for examining whether the understanding of certain affordances in AR persists, can be recalled, or changes over time.
Besides, expanding evaluation to real-world settings would also allow for comparisons between lab and field studies, testing the generalizability of findings from controlled environments.



\section{Conclusion}
\label{ch:conclusion}

In this position paper, we discussed fallacies associated with the use of affordances in AR.
For that, we described what affordance is, identified the problems we have encountered, and proposed ways to begin addressing them.
In general, the productive use of (perceived) affordance in future AR still requires time and effort.
To achieve this, we need more information through studies and evaluations that will enable us to make meaningful assumptions and decisions.
However, despite our critical stance on the current use of affordances opportunistically, we remain positive on the integration in AR applications -- especially since AR's core feature is to combine virtual and physical scenes~\cite{steffen2019framework}.
Affordances are an excellent tool for envisioning novel AR design paradigms, both for placement and interactions.
They allow us to explore the full potential of the environment and its objects, encouraging creativity and imaginative applications -- like being dreamers (\cf~\cite{schawel2014waltdisneymethode}).
Nevertheless, we want to emphasize the need for critical reflection and assessment of the techniques created through these explorations.
What value do novel interaction concepts provide if they never find practical application in the real world? 
With that, we advocate for a more "reflective" approach to affordances in AR applications.


\begin{acks}
This work was supported by the Deutsche Forschungsgemeinschaft (DFG, German Research Foundation) grant 389792660 as part of TRR~248 -- CPEC (see \url{https://cpec.science}).
\end{acks}

\bibliographystyle{ACM-Reference-Format}
\bibliography{bib}

\end{document}